\documentclass[aps,prd,twocolumn,preprintnumbers,nofootinbib,superscriptaddress,longbibliography]{revtex4-1}

\usepackage[utf8]{inputenc}
\usepackage{amsmath}
\usepackage{amssymb}
\usepackage{xspace}
\usepackage{xcolor}
\usepackage{color}
\usepackage{soul}
\usepackage{placeins} 
\usepackage{graphicx}
\usepackage{siunitx}
\usepackage[normalem]{ulem}
\usepackage[colorlinks=true,citecolor=blue,linkcolor=blue]{hyperref}
\usepackage[capitalise, english]{cleveref}
\crefformat{equation}{(#2#1#3)}
\usepackage{bm}

\definecolor{red}{rgb}{0.9, 0,0}
\definecolor{cerulean}{rgb}{0., 0.62,0.9}
\definecolor{navy}{rgb}{0.05, 0.05,0.8}

\graphicspath{{./figures/}}

\begin{document}

\title{Bounds on axion-like particles from fully merged photons at the LHC}

\author{Simon Knapen}
\email{smknapen@lbl.gov}
\affiliation{Theoretical Physics Group, Lawrence Berkeley National Laboratory, Berkeley, CA 94720, USA}
\affiliation{Leinweber Institute for Theoretical Physics, Department of Physics, University of California, \mbox{Berkeley}, CA 94720, USA}

\begin{abstract} 
 I show that, in the mass range $0.1\,\text{GeV} < m_a < 6\,\text{GeV}$, the strongest current bound on axion-like particles (ALPs) coupled to photons  comes from very high energy $a\gamma$ production at the LHC. In this regime, the two photons from the ALP decay are collimated and reconstructed as a single photon. 
\end{abstract}

\maketitle

\textbf{Introduction:} Extensions of the Standard Model (SM) could contain one or more light particles that couple predominantly to photons, similar to the $\eta$ and $\pi^0$ in the SM. The theoretical motivation for their presence in the LHC data is summarized in e.g.~\cite{CidVidal:2018blh,Hook:2019qoh,Knapen:2021elo,Gershtein:2020mwi}.
Below the electroweak scale, such an axion-like particle ($a$) interacts with the Standard Model  through the operator
\begin{equation}\label{eq:definition}
    \mathcal{L}\supset - \frac{1}{4} g_{a\gamma\gamma} a F\tilde F 
\end{equation}
with $\tilde F^{\mu\nu}\equiv \frac{1}{2}\epsilon^{\mu\nu\rho\sigma}F_{\rho\sigma}$.
Since this is a dimension five operator, the parton cross section at high energies is independent of the collision energy
\begin{align}\label{eq:signalsigma}
\hat\sigma_{q\bar q \to a\gamma} &= \frac{\alpha}{72}\, Q_q^2\, g_{a\gamma\gamma}^2
\end{align}
with $Q_q$ the electric charge of the quark and $\alpha$ the fine-structure constant. The dominant SM background is the $q\bar q\to \gamma\gamma$ process, assuming that the photons from the ALP decay are reconstructed as a single photon.
The corresponding parton cross section for the angular range $\theta_0<\theta<\pi-\theta_0$
is
\begin{align}\label{eq:background}
\sigma_{q\bar q \to \gamma\gamma} &= \frac{2\pi\alpha^2 Q_q^4}{3m_{\gamma\gamma}^2} \left[ \ln\frac{1+\cos\theta_0}{1-\cos\theta_0} - \cos\theta_0 \right].
\end{align}
Unlike \eqref{eq:signalsigma}, it drops quadratically with the diphoton invariant mass ($m_{\gamma\gamma}$).
Taking $\theta_0=10^\circ$ and $Q_q=2/3$, the signal-to-background ratio is then approximately
\begin{equation}\label{eq:ratio}
\frac{\sigma_{q\bar q \to a\gamma}}{\sigma_{q\bar q \to \gamma\gamma}}\approx 0.5 \times (m_{\gamma\gamma} g_{a\gamma\gamma})^2.
\end{equation}
Since ATLAS and CMS have seen diphoton pairs well above $m_{\gamma\gamma}> 2$ TeV \cite{ATLAS:2021uiz,CMS:2024nht}, the estimate in \eqref{eq:ratio} therefore suggests that the high energy tail of the $m_{\gamma\gamma}$ distribution can constrain the ALP-photon coupling down to $g_{a\gamma\gamma} \approx 4\times 10^{-4}\, \mathrm{GeV}^{-1}$.

\textbf{Analysis:} To carry out a faithful LHC analysis, one must consider the full electroweak theory that gives rise to \eqref{eq:definition}, which is
\begin{equation}
\mathcal{L}\supset-\frac{1}{4}\frac{c_B}{f}B\tilde B - \frac{1}{4}\frac{c_W}{f}W^i\tilde W^i
\end{equation}
with $B$ and $W^i$ the hypercharge and $SU(2)$ field strength tensors. In the mass basis, the $a$-$\gamma$-$\gamma$ and $a$-$\gamma$-$Z$ couplings are
\begin{align}
g_{a\gamma\gamma}=&\frac{1}{f}(\cos^2{\theta_w} c_B + \sin^2{\theta_w} c_W)\\
g_{a\gamma Z}=&\frac{1}{f}\sin{2\theta_w} (c_W-c_B)\label{eq:gammaZ}
\end{align}
with $\theta_W$ the Weinberg angle. I assume $c_B\gg c_W$, and comment on other choices in the results section.

\begin{figure}
    \centering
\includegraphics[width=\linewidth]{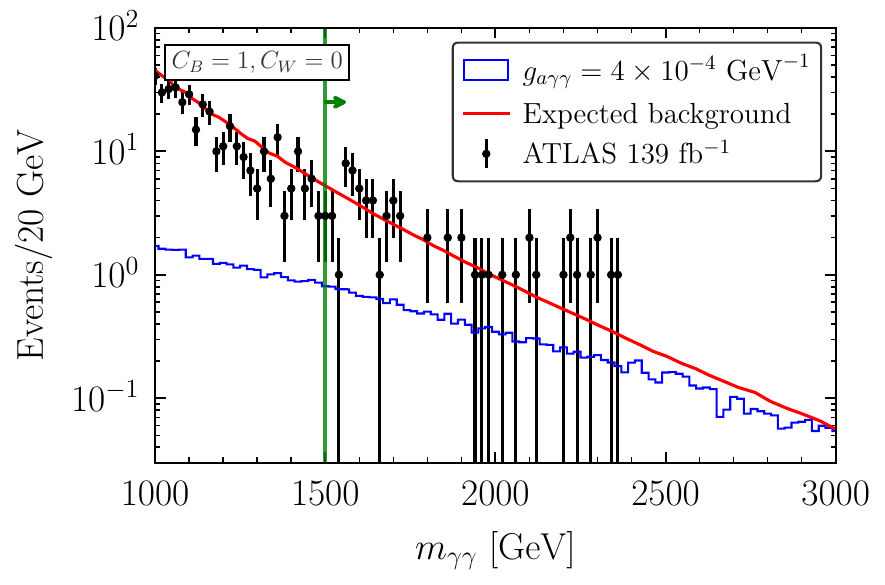}
    \caption{Expected backgrounds and observed events \cite{ATLAS:2021uiz} and the ALP signal for $m_a=1$ GeV, $g_{a\gamma\gamma}=4\times 10^{-4}\,\mathrm{GeV}^{-1}$, after all selection cuts. The data included in the analysis is to the right of the vertical green line. The errors are statistical only.}
    \label{fig:spectrum}
\end{figure}

\begin{figure*}[t!]
    \centering
\includegraphics[width=\linewidth]{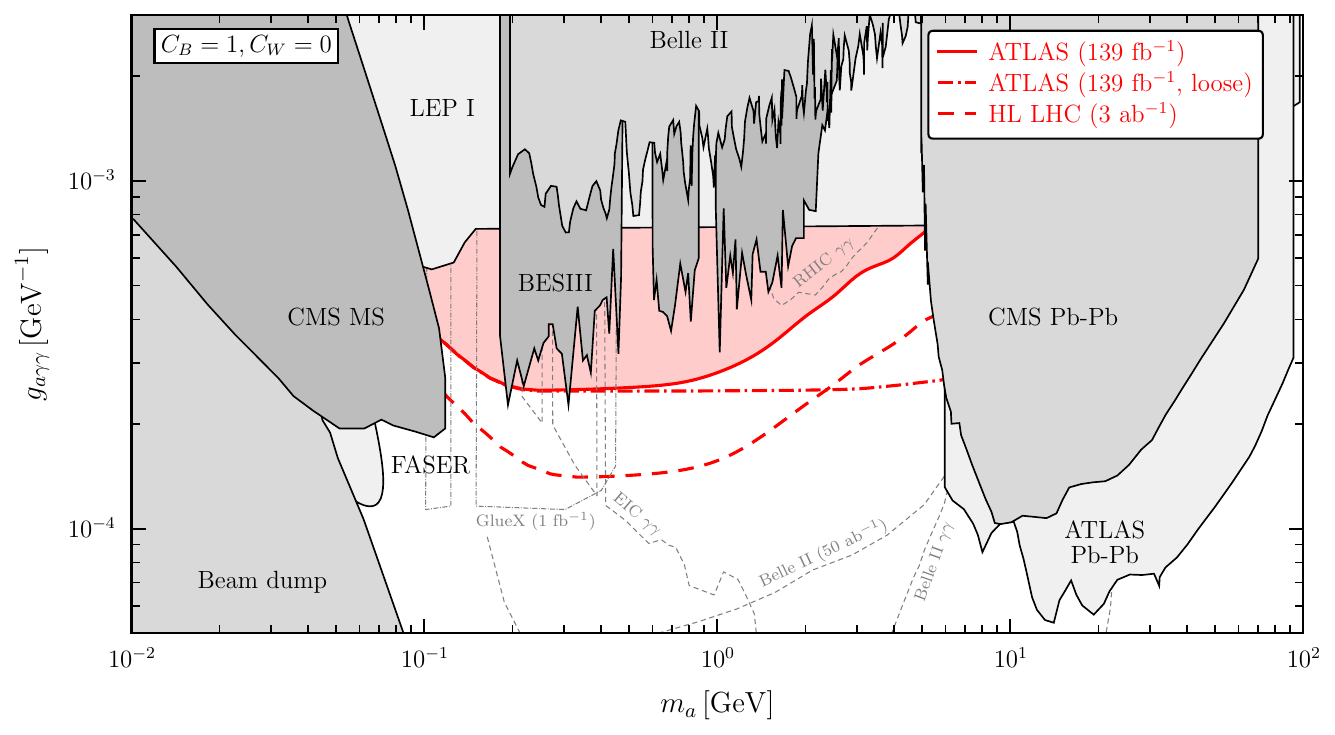}
    \caption{$95\%$ confidence bound derived from the ATLAS high mass $\gamma\gamma$ resonance search \cite{ATLAS:2021uiz} (solid red), along with an estimate of potential bound with loose photon selection (dot-dashed) and projection for the HL LHC (dashed). Existing bounds are shown in gray, including beam dumps (NuCal) \cite{Dobrich:2019dxc,Blumlein:1990ay}, FASER \cite{FASER:2024bbl}, LEP I \cite{Jaeckel:2015jla}, Belle II \cite{Belle-II:2020jti}, BESIII \cite{BESIII:2024hdv}, the CMS search for displaced decays in the muon chambers ("CMS MS") \cite{Mitridate:2023tbj,CMS:2021juv} and the ATLAS and CMS heavy ion runs \cite{ATLAS:2020hii,CMS:2018erd}(``ATLAS Pb-Pb'' and ``CMS Pb-Pb''). For clarity, some bounds were omitted if they are fully covered by another bound, e.g.~the ATLAS $Z\to 3\gamma$ analysis \cite{ATLAS:2015rsn}. 
    The dashed gray lines show projected bounds for GlueX \cite{Aloni:2019ruo,Pybus:2023yex}, Belle II using $e^+e^-\to a\gamma$ \cite{Dolan:2017osp}  and $\gamma\gamma$-fusion \cite{Acanfora:2024spi}, as well as $\gamma\gamma$-fusion at the electron-ion collider (EIC) \cite{Balkin:2023gya} and the relativistic heavy ion collider (RHIC) \cite{Fuyuto:2026tuc}. 
    }
    \label{fig:result}
\end{figure*}

To extract a bound on $g_{a\gamma\gamma}$, I have recast the ATLAS analysis for high mass diphoton resonances, which was performed with an integrated luminosity of 139 $\mathrm{fb}^{-1}$ \cite{ATLAS:2021uiz}. 
The fiducial selection of the search was $|\eta|<2.37$ for both photons and $p_T/m_{\gamma\gamma}>0.3$ ($p_T/m_{\gamma\gamma}>0.25$) for the first (second) photon, along with standard photon isolation requirements.
The events observed by ATLAS and expected background from $q\bar q\to \gamma\gamma$ are shown in \cref{fig:spectrum}.
ATLAS obtained this background model from a fit to Monte Carlo and an uncorrelated control sample; it is therefore not contaminated by the signal I consider.
I have generated signal events at leading order with \verb+Madgraph+ \cite{Alwall:2014hca} using the UFO model from \cite{Brivio:2017ije} with a k-factor of 1.3, obtained with \verb+Madgraph@NLO+ for the similar $pp\to \gamma\gamma$ process at high $p_T$.
I further placed the same selection cuts as the ATLAS analysis, treating the ALP as a single photon and require it to decay before the inner radius of the electromagnetic calorimeter (ECAL). 
To ensure my merged-photon assumption is justified, I also require that the photons from the ALP decay are close enough to be reconstructed as a single, tight photon (see Appendix~\ref{app:selectioneff}).
In addition, I applied the parametrized reconstruction efficiency provided in the supplementary material of \cite{ATLAS:2021uiz}. 
The signal spectrum after all these selections is shown by the blue histogram in \cref{fig:spectrum}.
The falling signal spectrum is entirely due to $q\bar q$ parton luminosity, while the background falls faster due to the additional $1/m_{\gamma\gamma}^2$ dependence in \cref{eq:background}.

To extract the 95\% confidence level bound, I have performed a profile-likelihood ratio test on all bins with $m_{\gamma\gamma}>1.5$ TeV using the \verb+pyhf+ package \cite{pyhf,pyhf_joss}, assuming negligible uncertainties on the background prediction (see Appendix~\ref{app:systematic}).
The limits were extracted using the asymptotics formulas and then validated with toy simulations. As an additional cross check, I have also calculated the limit for a single bin containing all $m_{\gamma\gamma}>1.5$ TeV events. 
The limit on the signal strength in this case was roughly 10\% weaker than in the binned analysis.
Overall the limit is slightly stronger than the expected limit due to a modest downward fluctuation of the data relative to the background model.

\textbf{Discussion:} The resulting bound is shown by the red shaded area in \cref{fig:result}.\footnote{The limits on this figure were in part taken from the AxionLimits repository \cite{AxionLimits}.}
The ATLAS data provides the strongest bound for most of the $0.1\,\text{GeV} \lesssim m_a \lesssim 6\,\text{GeV}$ region. The limit weakens for $m_a \lesssim 0.2$ GeV and $m_a\gtrsim 1$ GeV. The former is because the ALP decay is increasingly too displaced, while the latter  
is due to my conservative treatment of the tight photon selection criteria (see Appendix~\ref{app:selectioneff}). 
To illustrate the importance of this selection, I also show what the limit would have been with loose photons, assuming the background remains the same (dot-dashed line).
The latter is of course optimistic, but I suspect it to be a fairly good estimate, since the supplementary material to \cite{ATLAS:2021uiz} states that the tight photon selection only reduces the number of observed events by about 10\% in their $m_{\gamma\gamma}=1$ TeV bin. 
It seems plausible that a similar or smaller rescaling would hold at higher $m_{\gamma\gamma}$ as well.
I am nevertheless not claiming this line as a limit, but suggest it as an opportunity for ATLAS to improve upon this study by considering one loose and one tight photon.\footnote{A dedicated analysis that can resolve the substructure in the photon candidate~\cite{CMS:2024vjn,ATLAS:2018dfo}
 may further increase the sensitivity.} 
I also estimated the sensitivity for the HL LHC, by showing expected limit with a luminosity of $3\,\text{ab}^{-1}$.

It is worth noting that the Belle II bound \cite{Belle-II:2020jti} a priori seems  surprisingly weak. This is because only $445\,\mathrm{pb}^{-1}$ of luminosity was included. 
In the meanwhile Belle II has collected about $900\,\mathrm{fb}^{-1}$ and the collaboration aims to ultimately collect $50\,\mathrm{ab}^{-1}$. 
If it is not systematics limited, this ultimate data set is expected to outperform my HL LHC projection \cite{Dolan:2017osp}, especially if the $\gamma\gamma$-fusion process is included \cite{Acanfora:2024spi}.
Other than the Belle II prospects, I am aware of proposals to search in the same parameter space with ultra-peripheral heavy ion collisions at RHIC \cite{Fuyuto:2026tuc} and the EIC \cite{Balkin:2023gya} and through Primakoff production at GlueX \cite{Aloni:2019ruo,Pybus:2023yex}. 
For all current and future sensitivities, the UV completion of \cref{eq:definition} contains either a fairly large number of heavy charged particles or an smaller number of relatively light charged particles that have somehow evaded detection at LEP and/or the LHC. The ATLAS bound assumes the former and all such charged particles need to have a mass above the TeV scale.   

Finally, with the $c_W\ll c_B$ assumption, \cref{eq:gammaZ} predicts $g_{a\gamma Z}/g_{a\gamma\gamma}=-2\tan \theta_w\approx-1.1$. A similar constraint can therefore be derived from the searches for high mass $Z\gamma$ resonances \cite{ATLAS:2023wqy,CMS:2017dyb}. I however expect it to be weaker than the $\gamma\gamma$ case because of the $Z\to \ell\ell$ branching ratios that need to be included for the cleanest channels.
While I assumed $c_W\ll c_B$, the $c_W\approx c_B$ ansatz is also frequently used in the literature. I also calculated the ATLAS bound with this alternative assumption and find it to be slightly stronger.
The $c_W\approx c_B$ ansatz is popular because it suppresses the ALP-photon-Z coupling in \eqref{eq:gammaZ}, thus removing constraints from $Z\to a\gamma$ decays. In particular, the \mbox{LEP I} bound on $Z\to a\gamma$ \cite{Jaeckel:2015jla} decays doesn't apply and one must fall back on a significantly weaker bound from modifications to $e^+e^-\to \gamma\gamma$ cross section as measured by OPAL \cite{Knapen:2016moh,OPAL:2002vhf}. 
However, BaBar and Belle have since constrained $c_W/f$ through a search for $B\to K a$ decays \cite{BaBar:2021ich,Belle:2025zbq,Izaguirre:2016dfi}; for the $c_W\approx c_B$ scenario this corresponds to a very strong bound of $g_{a\gamma\gamma}\approx c_W/f\lesssim 5\times 10^{-6}\,\mathrm{GeV}^{-1}$. This is much stronger than the ATLAS bound I obtain here, or any other existing or projected direct bound on $g_{a\gamma\gamma}$ that I am aware of.

There are few extensions of this analysis that could of interest: In addition to the aforementioned $Z\gamma$ channel, it would be interesting to consider the high mass $\gamma$-jet resonance searches \cite{CMS:2023twl,ATLAS:2017dpx}, which apply to models where the ALP has a coupling to gluons. At low $m_a$ the ALP decays displaced, as was already considered in \cite{Mitridate:2023tbj}. In this regime I expect that the $\gamma$+missing momentum searches \cite{CMS:2018ffd,ATLAS:2020uiq} can provide a constraint, though likely only in a narrow $m_a$ range. I leave these cases for future work.



\section*{Acknowledgements}
I am thankful to the members of the Berkeley theory and ATLAS groups for comments and discussions, Dean Robinson and Simone Pagan Griso in particular. I also thank Diego Redigolo and Giulio Dainelli for insightful comments on the draft. This work is supported by the Office of High Energy Physics of the U.S.\ Department of Energy under contract DE-AC02-05CH11231. 

\bibliography{refs.bib}

@article{Brivio:2017ije,
    author = "Brivio, I. and Gavela, M. B. and Merlo, L. and Mimasu, K. and No, J. M. and del Rey, R. and Sanz, V.",
    title = "{ALPs Effective Field Theory and Collider Signatures}",
    eprint = "1701.05379",
    archivePrefix = "arXiv",
    primaryClass = "hep-ph",
    reportNumber = "IFT-UAM-CSIC-16-141, KCL-PH-TH-2016-72, FTUAM-16-49, CP3-17-04",
    doi = "10.1140/epjc/s10052-017-5111-3",
    journal = "Eur. Phys. J. C",
    volume = "77",
    number = "8",
    pages = "572",
    year = "2017"
}

@article{Belle:2025zbq,
    author = "Adachi, I. and others",
    collaboration = "Belle, Belle-II",
    title = "{Search for an Axion-Like Particle in $B\rightarrow K^{(*)} a (\rightarrow\gamma\gamma)$ Decays at Belle}",
    eprint = "2507.01249",
    archivePrefix = "arXiv",
    primaryClass = "hep-ex",
    reportNumber = "Belle II Preprint: 2025-017 KEK Preprint: 2025-16",
    doi = "10.1007/JHEP12(2025)109",
    journal = "JHEP",
    volume = "12",
    pages = "109",
    year = "2025"
}

@article{CMS:2024vjn,
    author = "Hayrapetyan, Aram and others",
    collaboration = "CMS",
    title = "{Search for New Resonances Decaying to Pairs of Merged Diphotons in Proton-Proton Collisions at s=13{\,}{\,}TeV}",
    eprint = "2405.00834",
    archivePrefix = "arXiv",
    primaryClass = "hep-ex",
    reportNumber = "CMS-EXO-22-022",
    doi = "10.1103/PhysRevLett.134.041801",
    journal = "Phys. Rev. Lett.",
    volume = "134",
    number = "4",
    pages = "041801",
    year = "2025"
}

@article{Catani:2018krb,
    author = "Catani, Stefano and Cieri, Leandro and de Florian, Daniel and Ferrera, Giancarlo and Grazzini, Massimiliano",
    title = "{Diphoton production at the LHC: a QCD study up to NNLO}",
    eprint = "1802.02095",
    archivePrefix = "arXiv",
    primaryClass = "hep-ph",
    reportNumber = "ZH-TH-06-18",
    doi = "10.1007/JHEP04(2018)142",
    journal = "JHEP",
    volume = "04",
    pages = "142",
    year = "2018"
}

@article{Blumlein:1990ay,
    author = "Blumlein, J. and others",
    title = "{Limits on neutral light scalar and pseudoscalar particles in a proton beam dump experiment}",
    reportNumber = "PHE-90-03",
    doi = "10.1007/BF01548556",
    journal = "Z. Phys. C",
    volume = "51",
    pages = "341--350",
    year = "1991"
}

@article{Dobrich:2019dxc,
    author = {D{\"o}brich, Babette and Jaeckel, Joerg and Spadaro, Tommaso},
    title = "{Light in the beam dump - ALP production from decay photons in proton beam-dumps}",
    eprint = "1904.02091",
    archivePrefix = "arXiv",
    primaryClass = "hep-ph",
    doi = "10.1007/JHEP05(2019)213",
    journal = "JHEP",
    volume = "05",
    pages = "213",
    year = "2019",
    note = "[Erratum: JHEP 10, 046 (2020)]"
}

@article{Pybus:2023yex,
    author = "Pybus, J. R. and others",
    title = "{Search for axion-like particles through nuclear Primakoff production using the GlueX detector}",
    eprint = "2308.06339",
    archivePrefix = "arXiv",
    primaryClass = "hep-ex",
    doi = "10.1016/j.physletb.2024.138790",
    journal = "Phys. Lett. B",
    volume = "855",
    pages = "138790",
    year = "2024"
}

@article{Balkin:2023gya,
    author = "Balkin, Reuven and Hen, Or and Li, Wenliang and Liu, Hongkai and Ma, Teng and Soreq, Yotam and Williams, Mike",
    title = "{Probing axion-like particles at the Electron-Ion Collider}",
    eprint = "2310.08827",
    archivePrefix = "arXiv",
    primaryClass = "hep-ph",
    doi = "10.1007/JHEP02(2024)123",
    journal = "JHEP",
    volume = "02",
    pages = "123",
    year = "2024"
}

@article{Acanfora:2024spi,
    author = "Acanfora, Francesca and Franceschini, Roberto and Mastroddi, Alessio and Redigolo, Diego",
    title = "{Fusing photons into diphoton resonances at Belle II and beyond}",
    eprint = "2406.14614",
    archivePrefix = "arXiv",
    primaryClass = "hep-ph",
    doi = "10.1007/JHEP12(2024)099",
    journal = "JHEP",
    volume = "12",
    pages = "099",
    year = "2024"
}

@article{NNPDF:2021njg,
    author = "Ball, Richard D. and others",
    collaboration = "NNPDF",
    title = "{The path to proton structure at 1{\%} accuracy}",
    eprint = "2109.02653",
    archivePrefix = "arXiv",
    primaryClass = "hep-ph",
    reportNumber = "Edinburgh 2021/12, Nikhef-2021-013, TIF-UNIMI-2021-11",
    doi = "10.1140/epjc/s10052-022-10328-7",
    journal = "Eur. Phys. J. C",
    volume = "82",
    number = "5",
    pages = "428",
    year = "2022"
}

@article{Buckley:2014ana,
    author = {Buckley, Andy and Ferrando, James and Lloyd, Stephen and Nordstr{\"o}m, Karl and Page, Ben and R{\"u}fenacht, Martin and Sch{\"o}nherr, Marek and Watt, Graeme},
    title = "{LHAPDF6: parton density access in the LHC precision era}",
    eprint = "1412.7420",
    archivePrefix = "arXiv",
    primaryClass = "hep-ph",
    reportNumber = "GLAS-PPE-2014-05, MCNET-14-29, IPPP-14-111, DCPT-14-222",
    doi = "10.1140/epjc/s10052-015-3318-8",
    journal = "Eur. Phys. J. C",
    volume = "75",
    pages = "132",
    year = "2015"
}

@article{PDF4LHCWorkingGroup:2022cjn,
    author = "Ball, Richard D. and others",
    collaboration = "PDF4LHC Working Group",
    title = "{The PDF4LHC21 combination of global PDF fits for the LHC Run III}",
    eprint = "2203.05506",
    archivePrefix = "arXiv",
    primaryClass = "hep-ph",
    reportNumber = "Edinburgh 2021/31, FERMILAB-PUB-22-121-QIS-SCD-T, MSUHEP-22-010, SMU-HEP-22-01, Nikhef 2021-033",
    doi = "10.1088/1361-6471/ac7216",
    journal = "J. Phys. G",
    volume = "49",
    number = "8",
    pages = "080501",
    year = "2022"
}

@article{ATLAS:2018dfo,
    author = "Aaboud, Morad and others",
    collaboration = "ATLAS",
    title = "{A search for pairs of highly collimated photon-jets in $pp$ collisions at $\sqrt{s}$ = 13 TeV with the ATLAS detector}",
    eprint = "1808.10515",
    archivePrefix = "arXiv",
    primaryClass = "hep-ex",
    reportNumber = "CERN-EP-2018-143",
    doi = "10.1103/PhysRevD.99.012008",
    journal = "Phys. Rev. D",
    volume = "99",
    number = "1",
    pages = "012008",
    year = "2019"
}

@article{CMS:2024nht,
    author = "Hayrapetyan, Aram and others",
    collaboration = "CMS",
    title = "{Search for new physics in high-mass diphoton events from proton-proton collisions at $ \sqrt{\textrm{s}} $ = 13 TeV}",
    eprint = "2405.09320",
    archivePrefix = "arXiv",
    primaryClass = "hep-ex",
    reportNumber = "CMS-EXO-22-024, CERN-EP-2024-109",
    doi = "10.1007/JHEP08(2024)215",
    journal = "JHEP",
    volume = "08",
    pages = "215",
    year = "2024"
}

@article{CidVidal:2018blh,
    author = "Cid Vidal, Xabier and Mariotti, Alberto and Redigolo, Diego and Sala, Filippo and Tobioka, Kohsaku",
    title = "{New Axion Searches at Flavor Factories}",
    eprint = "1810.09452",
    archivePrefix = "arXiv",
    primaryClass = "hep-ph",
    reportNumber = "DESY-18-183",
    doi = "10.1007/JHEP01(2019)113",
    journal = "JHEP",
    volume = "01",
    pages = "113",
    year = "2019",
    note = "[Erratum: JHEP 06, 141 (2020)]"
}

@article{Hook:2019qoh,
    author = "Hook, Anson and Kumar, Soubhik and Liu, Zhen and Sundrum, Raman",
    title = "{High Quality QCD Axion and the LHC}",
    eprint = "1911.12364",
    archivePrefix = "arXiv",
    primaryClass = "hep-ph",
    reportNumber = "UMD-PP-019-07",
    doi = "10.1103/PhysRevLett.124.221801",
    journal = "Phys. Rev. Lett.",
    volume = "124",
    number = "22",
    pages = "221801",
    year = "2020"
}

@article{Gershtein:2020mwi,
    author = "Gershtein, Yuri and Knapen, Simon and Redigolo, Diego",
    title = "{Probing naturally light singlets with a displaced vertex trigger}",
    eprint = "2012.07864",
    archivePrefix = "arXiv",
    primaryClass = "hep-ph",
    reportNumber = "CERN-TH-2020-207",
    doi = "10.1016/j.physletb.2021.136758",
    journal = "Phys. Lett. B",
    volume = "823",
    pages = "136758",
    year = "2021"
}

@article{Knapen:2021elo,
    author = "Knapen, Simon and Kumar, Soubhik and Redigolo, Diego",
    title = "{Searching for axionlike particles with data scouting at ATLAS and CMS}",
    eprint = "2112.07720",
    archivePrefix = "arXiv",
    primaryClass = "hep-ph",
    reportNumber = "CERN-TH-2021-216",
    doi = "10.1103/PhysRevD.105.115012",
    journal = "Phys. Rev. D",
    volume = "105",
    number = "11",
    pages = "115012",
    year = "2022"
}

@article{ATLAS:2017dpx,
    author = "Aaboud, Morad and others",
    collaboration = "ATLAS",
    title = "{Search for new phenomena in high-mass final states with a photon and a jet from $pp$ collisions at $\sqrt{s}$ = 13 TeV with the ATLAS detector}",
    eprint = "1709.10440",
    archivePrefix = "arXiv",
    primaryClass = "hep-ex",
    reportNumber = "CERN-EP-2017-148",
    doi = "10.1140/epjc/s10052-018-5553-2",
    journal = "Eur. Phys. J. C",
    volume = "78",
    number = "2",
    pages = "102",
    year = "2018"
}

@article{ATLAS:2020uiq,
    author = "Aad, Georges and others",
    collaboration = "ATLAS",
    title = "{Search for dark matter in association with an energetic photon in $pp$ collisions at $\sqrt{s}$ = 13 TeV with the ATLAS detector}",
    eprint = "2011.05259",
    archivePrefix = "arXiv",
    primaryClass = "hep-ex",
    reportNumber = "CERN-EP-2020-178",
    doi = "10.1007/JHEP02(2021)226",
    journal = "JHEP",
    volume = "02",
    pages = "226",
    year = "2021"
}

@article{CMS:2023twl,
    author = "Tumasyan, Armen and others",
    collaboration = "CMS",
    title = "{Search for resonances in events with photon and jet final states in proton-proton collisions at $ \sqrt{s} $ = 13 TeV}",
    eprint = "2305.07998",
    archivePrefix = "arXiv",
    primaryClass = "hep-ex",
    reportNumber = "CMS-EXO-20-012, CERN-EP-2023-064",
    doi = "10.1007/JHEP12(2023)189",
    journal = "JHEP",
    volume = "12",
    pages = "189",
    year = "2023"
}

@article{CMS:2018ffd,
    author = "Sirunyan, Albert M and others",
    collaboration = "CMS",
    title = "{Search for new physics in final states with a single photon and missing transverse momentum in proton-proton collisions at $\sqrt{s} =$ 13 TeV}",
    eprint = "1810.00196",
    archivePrefix = "arXiv",
    primaryClass = "hep-ex",
    reportNumber = "CMS-EXO-16-053, CERN-EP-2018-248",
    doi = "10.1007/JHEP02(2019)074",
    journal = "JHEP",
    volume = "02",
    pages = "074",
    year = "2019"
}

@article{CMS:2021juv,
    author = "Tumasyan, Armen and others",
    collaboration = "CMS",
    title = "{Search for Long-Lived Particles Decaying in the CMS End Cap Muon Detectors in Proton-Proton Collisions at $\sqrt s$ =13{\,}{\,}TeV}",
    eprint = "2107.04838",
    archivePrefix = "arXiv",
    primaryClass = "hep-ex",
    reportNumber = "CMS-EXO-20-015, CERN-EP-2021-125",
    doi = "10.1103/PhysRevLett.127.261804",
    journal = "Phys. Rev. Lett.",
    volume = "127",
    number = "26",
    pages = "261804",
    year = "2021"
}

@article{Mitridate:2023tbj,
    author = "Mitridate, Andrea and Papucci, Michele and Wang, Christina W. and Pe{\~n}a, Cristi{\'a}n and Xie, Si",
    title = "{Energetic long-lived particles in the CMS muon chambers}",
    eprint = "2304.06109",
    archivePrefix = "arXiv",
    primaryClass = "hep-ph",
    reportNumber = "CALT-TH-2023-001, DESY-2023-01670, FERMILAB-PUB-23-204-PPD",
    doi = "10.1103/PhysRevD.108.055040",
    journal = "Phys. Rev. D",
    volume = "108",
    number = "5",
    pages = "055040",
    year = "2023"
}

@article{Jaeckel:2015jla,
    author = "Jaeckel, Joerg and Spannowsky, Michael",
    title = "{Probing MeV to 90 GeV axion-like particles with LEP and LHC}",
    eprint = "1509.00476",
    archivePrefix = "arXiv",
    primaryClass = "hep-ph",
    doi = "10.1016/j.physletb.2015.12.037",
    journal = "Phys. Lett. B",
    volume = "753",
    pages = "482--487",
    year = "2016"
}

@article{Alwall:2014hca,
  author        = {Alwall, J. and Frederix, R. and Frixione, S. and Hirschi, V. and Maltoni, F. and Mattelaer, O. and Shao, H. -S. and Stelzer, T. and Torrielli, P. and Zaro, M.},
  title         = {{The automated computation of tree-level and next-to-leading order differential cross sections, and their matching to parton shower simulations}},
  eprint        = {1405.0301},
  archiveprefix = {arXiv},
  primaryclass  = {hep-ph},
  reportnumber  = {CERN-PH-TH-2014-064, CP3-14-18, LPN14-066, MCNET-14-09, ZU-TH-14-14},
  doi           = {10.1007/JHEP07(2014)079},
  journal       = {JHEP},
  volume        = {07},
  pages         = {079},
  year          = {2014}
}

@article{ATLAS:2018fzd,
    author = "Aaboud, Morad and others",
    collaboration = "ATLAS",
    title = "{Measurement of the photon identification efficiencies with the ATLAS detector using LHC Run 2 data collected in 2015 and 2016}",
    eprint = "1810.05087",
    archivePrefix = "arXiv",
    primaryClass = "hep-ex",
    reportNumber = "CERN-EP-2018-216",
    doi = "10.1140/epjc/s10052-019-6650-6",
    journal = "Eur. Phys. J. C",
    volume = "79",
    number = "3",
    pages = "205",
    year = "2019"
}

@article{FASER:2024bbl,
    author = "Mammen Abraham, Roshan and others",
    collaboration = "FASER",
    title = "{Shining light on the dark sector: search for axion-like particles and other new physics in photonic final states with FASER}",
    eprint = "2410.10363",
    archivePrefix = "arXiv",
    primaryClass = "hep-ex",
    reportNumber = "CERN-EP-2024-262",
    doi = "10.1007/JHEP01(2025)199",
    journal = "JHEP",
    volume = "01",
    pages = "199",
    year = "2025"
}

@article{OPAL:2002vhf,
    author = "Abbiendi, G. and others",
    collaboration = "OPAL",
    title = "{Multiphoton production in e+ e- collisions at s**(1/2) = 181-GeV to 209-GeV}",
    eprint = "hep-ex/0210016",
    archivePrefix = "arXiv",
    reportNumber = "CERN-EP-2002-060",
    doi = "10.1140/epjc/s2002-01074-5",
    journal = "Eur. Phys. J. C",
    volume = "26",
    pages = "331--344",
    year = "2003"
}

@article{CMS:2017dyb,
    author = "Sirunyan, Albert M and others",
    collaboration = "CMS",
    title = "{Search for Z$\gamma$ resonances using leptonic and hadronic final states in proton-proton collisions at $\sqrt{s}=$ 13 TeV}",
    eprint = "1712.03143",
    archivePrefix = "arXiv",
    primaryClass = "hep-ex",
    reportNumber = "CMS-EXO-17-005, CERN-EP-2017-301",
    doi = "10.1007/JHEP09(2018)148",
    journal = "JHEP",
    volume = "09",
    pages = "148",
    year = "2018"
}

@article{ATLAS:2023wqy,
    author = "Aad, Georges and others",
    collaboration = "ATLAS",
    title = "{Search for the Z{\ensuremath{\gamma}} decay mode of new high-mass resonances in pp collisions at s=13 TeV with the ATLAS detector}",
    eprint = "2309.04364",
    archivePrefix = "arXiv",
    primaryClass = "hep-ex",
    reportNumber = "CERN-EP-2023-153",
    doi = "10.1016/j.physletb.2023.138394",
    journal = "Phys. Lett. B",
    volume = "848",
    pages = "138394",
    year = "2024"
}

@article{Fuyuto:2026tuc,
    author = "Fuyuto, Kaori and Manzari, Claudio Andrea and Murayama, Hitoshi",
    title = "{Searches for GeV-Scale ALPs at RHIC}",
    eprint = "2606.07739",
    archivePrefix = "arXiv",
    primaryClass = "hep-ph",
    reportNumber = "RIKEN-iTHEMS-Report-26",
    month = "6",
    year = "2026"
}

@article{Aloni:2019ruo,
    author = "Aloni, Daniel and Fanelli, Cristiano and Soreq, Yotam and Williams, Mike",
    title = "{Photoproduction of Axionlike Particles}",
    eprint = "1903.03586",
    archivePrefix = "arXiv",
    primaryClass = "hep-ph",
    reportNumber = "CERN-TH-2019-023",
    doi = "10.1103/PhysRevLett.123.071801",
    journal = "Phys. Rev. Lett.",
    volume = "123",
    number = "7",
    pages = "071801",
    year = "2019"
}

@article{BESIII:2024hdv,
    author = "Ablikim, Medina and others",
    collaboration = "BESIII",
    title = "{Search for diphoton decays of an axionlike particle in radiative J/\ensuremath{\psi} decays}",
    eprint = "2404.04640",
    archivePrefix = "arXiv",
    primaryClass = "hep-ex",
    reportNumber = "BESIII Analysis Memo - 671",
    doi = "10.1103/PhysRevD.110.L031101",
    journal = "Phys. Rev. D",
    volume = "110",
    number = "3",
    pages = "L031101",
    year = "2024"
}

@article{CMS:2018erd,
    author = "Sirunyan, Albert M and others",
    collaboration = "CMS",
    title = "{Evidence for light-by-light scattering and searches for axion-like particles in ultraperipheral PbPb collisions at $\sqrt{s_\mathrm{NN}} =$ 5.02 TeV}",
    eprint = "1810.04602",
    archivePrefix = "arXiv",
    primaryClass = "hep-ex",
    reportNumber = "CMS-FSQ-16-012, CERN-EP-2018-271",
    doi = "10.1016/j.physletb.2019.134826",
    journal = "Phys. Lett. B",
    volume = "797",
    pages = "134826",
    year = "2019"
}

@article{ATLAS:2020hii,
    author = "Aad, Georges and others",
    collaboration = "ATLAS",
    title = "{Measurement of light-by-light scattering and search for axion-like particles with 2.2 nb$^{-1}$ of Pb+Pb data with the ATLAS detector}",
    eprint = "2008.05355",
    archivePrefix = "arXiv",
    primaryClass = "hep-ex",
    reportNumber = "CERN-EP-2020-135",
    doi = "10.1007/JHEP03(2021)243",
    journal = "JHEP",
    volume = "03",
    pages = "243",
    year = "2021",
    note = "[Erratum: JHEP 11, 050 (2021)]"
}

@software{pyhf,
  author = {Lukas Heinrich and Matthew Feickert and Giordon Stark},
  title = "{pyhf: v0.7.6}",
  version = {0.7.6},
  doi = {10.5281/zenodo.1169739},
  url = {https://doi.org/10.5281/zenodo.1169739},
  note = {https://github.com/scikit-hep/pyhf/releases/tag/v0.7.6}
}

@article{pyhf_joss,
  doi = {10.21105/joss.02823},
  url = {https://doi.org/10.21105/joss.02823},
  year = {2021},
  publisher = {The Open Journal},
  volume = {6},
  number = {58},
  pages = {2823},
  author = {Lukas Heinrich and Matthew Feickert and Giordon Stark and Kyle Cranmer},
  title = {pyhf: pure-Python implementation of HistFactory statistical models},
  journal = {Journal of Open Source Software}
}

@article{BaBar:2021ich,
    author = "Lees, J. P. and others",
    collaboration = "BaBar",
    title = "{Search for an Axionlike Particle in $B$ Meson Decays}",
    eprint = "2111.01800",
    archivePrefix = "arXiv",
    primaryClass = "hep-ex",
    reportNumber = "BABAR-PUB-21/006, SLAC-PUB-17631",
    doi = "10.1103/PhysRevLett.128.131802",
    journal = "Phys. Rev. Lett.",
    volume = "128",
    number = "13",
    pages = "131802",
    year = "2022"
}

@article{ATLAS:2015rsn,
    author = "Aad, Georges and others",
    collaboration = "ATLAS",
    title = "{Search for new phenomena in events with at least three photons collected in $pp$ collisions at $\sqrt{s}$ = 8 TeV with the ATLAS detector}",
    eprint = "1509.05051",
    archivePrefix = "arXiv",
    primaryClass = "hep-ex",
    reportNumber = "CERN-PH-EP-2015-187",
    doi = "10.1140/epjc/s10052-016-4034-8",
    journal = "Eur. Phys. J. C",
    volume = "76",
    number = "4",
    pages = "210",
    year = "2016"
}

@article{Belle-II:2020jti,
    author = "Abudin{\'e}n, F. and others",
    collaboration = "Belle-II",
    title = "{Search for Axion-Like Particles produced in $e^+e^-$ collisions at Belle II}",
    eprint = "2007.13071",
    archivePrefix = "arXiv",
    primaryClass = "hep-ex",
    reportNumber = "Belle II Preprint 2020-001, KEK Preprint 2020-10",
    doi = "10.1103/PhysRevLett.125.161806",
    journal = "Phys. Rev. Lett.",
    volume = "125",
    number = "16",
    pages = "161806",
    year = "2020"
}

@article{Izaguirre:2016dfi,
    author = "Izaguirre, Eder and Lin, Tongyan and Shuve, Brian",
    title = "{Searching for Axionlike Particles in Flavor-Changing Neutral Current Processes}",
    eprint = "1611.09355",
    archivePrefix = "arXiv",
    primaryClass = "hep-ph",
    reportNumber = "SLAC-PUB-16876",
    doi = "10.1103/PhysRevLett.118.111802",
    journal = "Phys. Rev. Lett.",
    volume = "118",
    number = "11",
    pages = "111802",
    year = "2017"
}

@article{Knapen:2016moh,
    author = "Knapen, Simon and Lin, Tongyan and Lou, Hou Keong and Melia, Tom",
    title = "{Searching for Axionlike Particles with Ultraperipheral Heavy-Ion Collisions}",
    eprint = "1607.06083",
    archivePrefix = "arXiv",
    primaryClass = "hep-ph",
    doi = "10.1103/PhysRevLett.118.171801",
    journal = "Phys. Rev. Lett.",
    volume = "118",
    number = "17",
    pages = "171801",
    year = "2017"
}

@article{Dolan:2017osp,
    author = "Dolan, Matthew J. and Ferber, Torben and Hearty, Christopher and Kahlhoefer, Felix and Schmidt-Hoberg, Kai",
    title = "{Revised constraints and Belle II sensitivity for visible and invisible axion-like particles}",
    eprint = "1709.00009",
    archivePrefix = "arXiv",
    primaryClass = "hep-ph",
    reportNumber = "DESY-17-127",
    doi = "10.1007/JHEP12(2017)094",
    journal = "JHEP",
    volume = "12",
    pages = "094",
    year = "2017",
    note = "[Erratum: JHEP 03, 190 (2021)]"
}

@misc{AxionLimits,
  author       = {Ciaran O'Hare},
  title        = {cajohare/AxionLimits: AxionLimits},
  month        = jul,
  year         = 2020,
  publisher    = {Zenodo},
  version      = {v1.0},
  doi          = {10.5281/zenodo.3932430},
  howpublished = {\url{https://cajohare.github.io/AxionLimits/}}
}

@article{ATLAS:2021uiz,
    author = "Aad, Georges and others",
    collaboration = "ATLAS",
    title = "{Search for resonances decaying into photon pairs in 139 fb$^{-1}$ of $pp$ collisions at $\sqrt {s}$=13 TeV with the ATLAS detector}",
    eprint = "2102.13405",
    archivePrefix = "arXiv",
    primaryClass = "hep-ex",
    reportNumber = "CERN-EP-2020-248",
    doi = "10.1016/j.physletb.2021.136651",
    journal = "Phys. Lett. B",
    volume = "822",
    pages = "136651",
    year = "2021"
}
\FloatBarrier
\appendix
\section{Details on selection efficiency}
\label{app:selectioneff}

Throughout this study I have assumed that the ALP is reconstructed as a single photon in the ATLAS detector. I use cuts on the $\Delta \phi$ and $\Delta\eta$ between both photons to enforce this condition on the signal events.

\begin{figure}
    \centering
    \includegraphics[width=\linewidth]{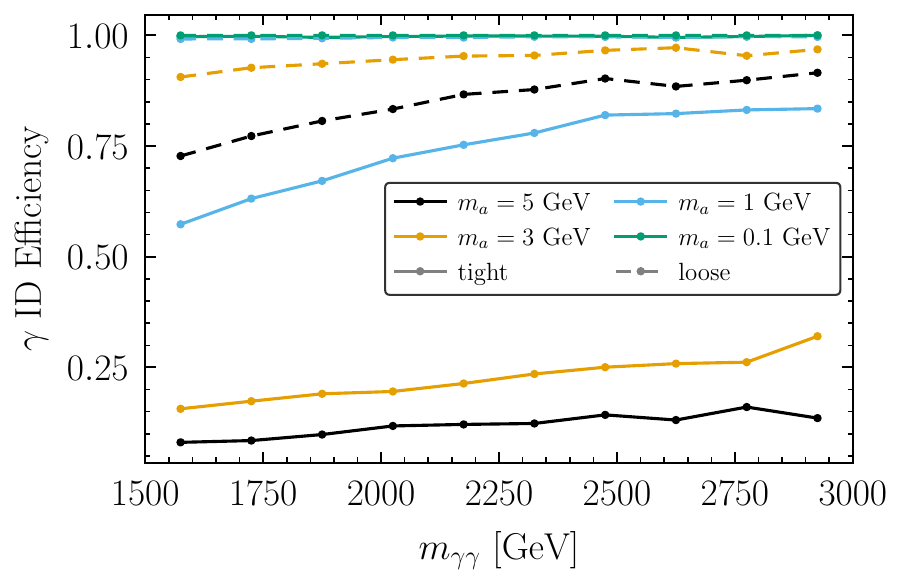}
    \caption{Signal efficiency for tight and loose photon selection, for ALPs that decayed before reaching the ECAL. See text for details.}
    \label{fig:efficiency}
\end{figure}

The main (second) layer of the ATLAS electromagnetic calorimeter has spatial granularity of approximately $0.025\times 0.025$ in $\eta\times\phi$ \cite{ATLAS:2018fzd}.
To identify photon candidates, cells are grouped in 3 $\times$ 5 grids of calorimeter cells, which are then processed by a clustering algorithm. 
For an ALP to be reconstructed as a single photon, I require both photons to land in a single cell or neighboring cells ($\Delta \eta<0.025$ and $\Delta \phi<0.025$).
The efficiency of this cut depends on $m_a$ and $m_{\gamma\gamma}$, as shown in the dashed lines in \cref{fig:efficiency}.

The first layer of the calorimeter however has a finer granularity in $\eta$, down to $0.003$ in some areas. 
The purpose of this layer is to reject showers that have a two-prong shape, which usually implies the origin was actually two collimated photons from a neutral hadron decay, e.g.~$\pi^0, \eta$ etc.
This is one of the criteria in the ``tight'' photon identification, which is what was used in \cite{ATLAS:2021uiz}.
``Loose'' photons on the other hand do not use information from the first layer of the calorimeter, and are therefore more prone to fakes from hadronic activity \cite{ATLAS:2018fzd}.

The exact selection for the tight photon identification is very complicated and depends among other things on the location in the calorimeter etc. 
It seems very unwise for a theorist to attempt to reproduce it. Instead, for the tight selection I demand that both photons satisfy $\Delta\eta<0.003$, such that they land in a single cell or neighboring cells of the first layer of the EM calorimeter.
The effect of this cut is shown by the solid line in \cref{fig:efficiency}.
Unfortunately, it is a rather costly cut for $m_a\gtrsim 1$ GeV.
It is plausible that this assumption is too conservative, and that some signal events could be recovered with the full ATLAS detector simulation.
Alternatively, ATLAS could search for an excess of events in the high $m_{\gamma\gamma}$ tail with one tight and one loose photon. 
The dot-dashed line in \cref{fig:result} is meant to represent an optimistic estimate of the latter scenario. 

\section{Systematic uncertainty estimate\label{app:systematic}}
I have neglected systematic uncertainties from the parton distribution functions (PDF's) in my limit setting procedure. The intuitive justification for this approximation is as follows: For $m_{\gamma\gamma}\geq 1.5$ TeV, the number of expected background events is 79.8, which corresponds to a statistical uncertainty on the background of 11\%. Bin-by-bin the statistical uncertainty is still higher and $\mathcal{O}(1)$ for the bins further on the tail of the distribution.
The ATLAS analysis estimates their systematic uncertainty on the background prediction to be between 10\% and 20\% of the statistical uncertainty and it thus has little impact on their limit. 
Their calculations are however tailored to a bump hunt and I therefore do not rely on them. 
The analogous CMS search \cite{CMS:2024nht} did place a limit on a continuum signal, and quoted $\sim5$\% for the systematic uncertainty on the PDFs. A $\sim5\%$ uncertainty is also consistent with the recent average for the $q\bar q$ luminosity \cite{PDF4LHCWorkingGroup:2022cjn}. 
In addition, both signal and background are proportional to the $q\bar q$ luminosity: For $m_{\gamma\gamma}>1$ TeV, ATLAS has bounded the contribution from jet+$\gamma$ processes to be $\lesssim 4$\% of the total background \cite{ATLAS:2021uiz}. The inclusive $\gamma\gamma$ spectrum does receive NLO contributions that are proportional to the $gq$ and $gg$ luminosities, however they are around $\sim 10$\% - $20$\% of the total $\gamma\gamma$ rate for the range of $m_{\gamma\gamma}$ that is of interest here \cite{Catani:2018krb}.
This means that the signal and the bulk of the background are proportional to same $q\bar q$ luminosity and the impact of systematic deviations in these PDFs is reduced.

\begin{figure}
    \centering
    \includegraphics[width=\linewidth]{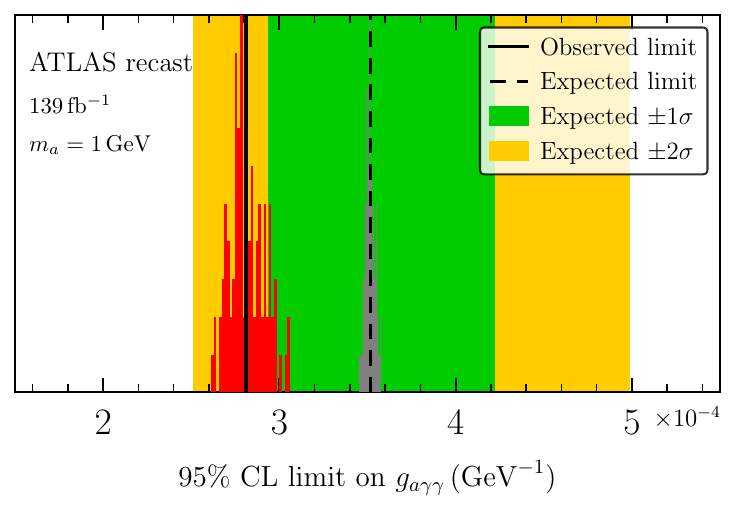}
    \caption{Histograms of expected (gray) and observed (red) limit for $m_a=1$ GeV over the NNPDF4.0 NNLO replica set \cite{NNPDF:2021njg}. The vertical dashed and solid line indicate the limit for the central value. The Brazil bands indicate the statistical uncertainty on the expected limit.}
    \label{fig:replicas}
\end{figure}

To make this intuition more rigorous, I recomputed the limit for the example point $m_a=1$ GeV using the NNPDF4.0 NNLO replica set \cite{NNPDF:2021njg} and the \verb+LHAPDF+ python interface \cite{Buckley:2014ana}: For each of the 100 replicas, the $q\bar q$ parton luminosity was computed, summing over the light quark flavors $q \in \{d,u,s,c,b\}$. I weighted each flavor by its electric charge squared $e_q^2$ to reflect the photon-induced production. 
The resulting per-replica luminosity curves provide an ensemble that carries the full bin-to-bin correlation structure of the PDF uncertainty.
Because both signal and background are proportional to the $q\bar q$ luminosity, I rescaled the signal and background yield in each mass bin by the ratio of the replica's luminosity to the central-set luminosity at that mass, and recomputed the limit.
This leaves the signal-to-background ratio within each bin unchanged by a given replica, and the residual PDF sensitivity arises only from the redistribution of statistical weight across mass bins. A partial cancellation of the PDF uncertainty in the limit is therefore expected, and reflects the fact that signal and background share the same $q\bar q$ initial-state luminosity.
This procedure yields a distribution of both expected and observed limits over the replica ensemble, as plotted in \cref{fig:replicas}.

The spread of the \emph{expected} limits across replicas (gray histogram) isolates the intrinsic effect of the PDF uncertainty on the search sensitivity, since the corresponding Asimov dataset varies together with each replica's background prediction and is therefore free of statistical fluctuations. This spread was found to be small as expected, specifically $\sigma_{\mathrm{PDF}} \approx 0.05 \times 10^{-4}\,\mathrm{GeV}^{-1}$, to be compared with the statistical uncertainty on the expected limit $\sigma_{\mathrm{STAT}}\approx 0.57 \times 10^{-4}\,\mathrm{GeV}^{-1}$. The PDF-induced variation is thus approximately $\sim10$\% of the statistical uncertainty and is therefore subdominant.

\cref{fig:replicas} also shows explicitly that the observed limit is about $\sim 1\sigma$ stronger than the expected limit, due to a mild downward fluctuation in the data relative to the expected background.
The spread on the observed limit (red histogram) due to the PDFs is larger than on the expected limit, but still comfortably smaller than the statistical uncertainty. 
This additional spread reflects the interplay between the mild downward fluctuation of the data and the replica-to-replica variation of the background shape, rather than an additional PDF uncertainty, and the expected-limit spread is therefore taken as the estimate of the systematic uncertainty. 

Other systematic uncertainties related to the trigger, photon isolation, photon identification, luminosity and pile-up are at the few \% level or smaller and small compared to the systematic uncertainty on the PDFs \cite{ATLAS:2021uiz}. 
I did not use the full ATLAS detector simulation and photon reconstruction algorithm to verify whether the photons from the ALP decay are reconstructed as a single photon. This introduces an additional uncertainty; however I have been maximally conservative by requiring both photons to be closer than the size of ECAL cells in the most granular layer of the ECAL.  
I therefore expect that a proper treatment by the ATLAS collaboration would result in a stronger limit than what I have presented here.

\end{document}